\documentclass[11pt,twoside]{article}
\usepackage{./asp2014}
\usepackage[para]{footmisc}

\aspSuppressVolSlug
\resetcounters

\begin{document}

\title{The First Six Outbursting Cool DA White Dwarf Pulsators}
\author{Keaton~J.~Bell,$^1$ J.~J.~Hermes,$^2$ M.~H.~Montgomery,$^1$ D.~E.~Winget,$^1$ N.~P.~Gentile~Fusillo,$^3$ R.~Raddi,$^3$ and B.~T.~G\"{a}nsicke$^3$
\affil{$^1$University of Texas at Austin, Austin, TX, USA; \email{keatonb@astro.as.utexas.edu}}
\affil{$^2$University of North Carolina, Chapel Hill, NC, USA; Hubble Fellow}
\affil{$^3$University of Warwick, Coventry, UK}}

\paperauthor{Keaton J. Bell}{keatonb@astro.as.utexas.edu}{0000-0002-0656-032X}{University of Texas at Austin}{Department of Astronomy}{Austin}{TX}{78712}{USA}
\paperauthor{J. J. Hermes}{jjhermes@unc.edu}{}{University of North Carolina}{Department of Physics and Astronomy}{Chapel Hill}{NC}{27599}{USA}
\paperauthor{M. H. Montgomery}{mikemon@astro.as.utexas.edu}{}{University of Texas at Austin}{Department of Astronomy}{Austin}{TX}{78712}{USA}
\paperauthor{D. E. Winget}{dew@astro.as.utexas.edu}{}{University of Texas at Austin}{Department of Astronomy}{Austin}{TX}{78712}{USA}
\paperauthor{N. P. Gentile Fusillo}{N.P.Gentile-Fusillo@warwick.ac.uk}{}{University of Warwick}{Department of Physics}{Coventry}{}{CV4 7AL}{UK}
\paperauthor{R. Raddi}{R.Raddi@warwick.ac.uk}{}{University of Warwick}{Department of Physics}{Coventry}{}{CV4 7AL}{UK}
\paperauthor{B. T. G\"{a}nsicke}{Boris.Gaensicke@warwick.ac.uk}{}{University of Warwick}{Department of Physics}{Coventry}{}{CV4 7AL}{UK}

\begin{abstract}
Extensive observations from the \emph{Kepler} spacecraft have recently revealed a new outburst phenomenon operating in cool pulsating DA (hydrogen atmosphere) white dwarfs (DAVs). With the introduction of two new outbursting DAVs from \emph{K2} Fields 7 (EPIC\,229228364) and 8 (EPIC\,220453225) in these proceedings, we presently know of six total members of this class of object. We present the observational commonalities of the outbursting DAVs: (1) outbursts that increase the mean stellar flux by up to $\approx$15\%, last many hours, and recur irregularly on timescales of days; (2) effective temperatures that locate them near the cool edge of the DAV instability strip; and (3) rich pulsation spectra with modes that are observed to wander in amplitude/frequency.
\end{abstract}

\section{Introduction}

Over 97\% of stars in the Milky Way ultimately end their lives as compact white dwarf stars.  Without appreciable nuclear fusion, white dwarfs evolve further by cooling.  When hydrogen-atmosphere (DA) white dwarfs near the average surface gravity of $\log{g}\approx 8$ cool through the range $12{,}500$\,K $> {T}_{\mathrm{eff}} > 10{,}600$\,K, partial ionization of atmospheric hydrogen induces stellar pulsations. We observe these DAs as photometric variables (DAVs), and the measured frequencies of variability are eigenfrequencies of the stars.  The tools of asteroseismology enable us to constrain the interior structures of DAVs from these measurements.

The \emph{Kepler} spacecraft observed one field of view nearly continuously for over 4 years in its original mission.  For a maximum of 512 pre-selected targets, time series photometry was collected at short cadence---roughly every 1\,min rather than every 30\,min.  With pulsation periods of $\sim 10$ minutes, short cadence \emph{Kepler} observations promised to capture by far the most complete record of DAV behavior.

\citet{Hermes2011} identified the first DAV in the original \emph{Kepler} mission field. WD\,J1916+3938 was observed by \emph{Kepler} at short cadence as KIC\,4552982 for over 1.5 years with a 86\% duty cycle.  Besides revealing a rich pulsation spectrum with 20 significant modes, these data captured a new outburst-like phenomenon operating in this star \citep{Bell2015}. A total of 178 flux enhancements reaching peaks of 2--17\% above the quiescent value and lasting 4--25\,hr were detected.  These outbursts carry a total energy of order $10^{33}$\,erg and have an average recurrence timescale of 2.7 days.  The observed time distribution of the outbursts favors delays longer than 2 days, beyond which their occurrences are consistent with Poisson statistics.  With a spectroscopic ${T}_{\mathrm{eff}} = 10{,}860 \pm 120$\,K at $\log{g} = 8.16\pm 0.06$, KIC\,4552982 is one of the coolest DAVs known.

After the \emph{Kepler} spacecraft's second reaction wheel failure in May 2013, the new \emph{K2} mission was devised for continued science operations in new fields in the ecliptic plane every $\approx$80 days \citep{Howell2014}.  \citet{Hermes2015} discovered another cool DAV---PG\,1149+057 (EPIC\,201806008) with ${T}_{\mathrm{eff}} = 11{,}060 \pm 170$ and $\log{g} = 8.06\pm 0.05$---to exhibit 10 outbursts in 78.8 days.  These outbursts caused instantaneous flux enhancements at high as 45\% that lasted between 9--36\,hr.  With a $Kepler$ magnitude of $K_p = 15.0$, this is the brightest outbursting DAV, and the corresponding high signal-to-noise ratio of the light curve enabled them to establish that the outbursts affect the pulsations---with pulsations generally having higher amplitudes and shorter periods during outbursts---definitively proving that these outbursts are occurring on the pulsating target star.

\citet{Bell2016} inspected the light curves of over 300 spectroscopically confirmed DA white dwarfs that were submitted for \emph{Kepler} observations through \emph{K2} Campaign 6 and discovered two additional outbursters in Fields 5 and 6: EPIC\,211629697 and EPIC\,229227292.  Both are also cool DAVs, demonstrating that only those within 500\,K of the empirical cool edge of the DAV instability strip are observed to undergo outbursts.  EPIC 211629697 ($\log{g} = 7.94\pm 0.08$, ${T}_{\mathrm{eff}} = 10{,}780 \pm 140$) showed 15 outbursts in 74.8 days of Campaign 5 data, with an average spacing of 5.0 days. These outbursts reached peaks of 8--15\% and lasted 6--38 hours.  EPIC\,229227292 ($\log{g} = 8.02\pm 0.05$, ${T}_{\mathrm{eff}} = 11{,}191 \pm 170$) is the most frequent outburster, exhibiting 33 outbursts every 2.4 days on average over 78.9 days of Campaign 6 observations. These reached peak fluxes of 4--9\% with 3--21 hour durations.

Alongside these observational developments, a possible physical mechanism has been proposed.  J.~J.\ Hermes described the potential for nonlinear mode coupling to cause outbursts in his talk at this conference, borrowing from the theoretical work of \citet{Wu2001}.  In this model, a resonant coupling can transfer energy from a driven parent mode into two daughter modes.  If these daughter modes are damped at the base of the convection zone, they will deposit their energy there, heating the surface of the star.  This mechanism for dumping pulsational energy may explain the empirical location of the cool edge of the DAV instability strip, which theoretical calculations predict to be thousands of degrees cooler than observed \citep[e.g.,][]{VanGrootel2012}.

In these proceedings, we present two more outbursting DAVs: EPIC\,229228364 and EPIC\,220453225, from \emph{K2} Fields 7 and 8.  We analyze the short cadence \emph{K2} data for these targets in Section~\ref{sec:anal}. We take a brief comparative look at the observational properties of the six known members of the new outbursting class of DAV in Section~\ref{sec:class}.  Finally, we discuss future observational prospects for these objects in Section~\ref{sec:future}.

\section{Two New Cool Outbursting DAVs}
\label{sec:anal}

We have discovered outbursts in the light curves of two new DAVs that we targeted for short cadence \emph{K2} observations based on their colors and proper motions.

EPIC\,229228364 ($K_p = 17.9$\,mag) was observed for 81.4 days in \emph{K2} Campaign 7.  The automated outburst detection algorithm described in \citet{Bell2016} found 6 significant outbursts in the light curve.  We measure the average outburst recurrence timescale to be 9.7 days.  These outbursts reach peak fluxes between 9--15\% above quiescence, and sustain their flux enhancement for 11--16 hours.

EPIC\,220453225 ($K_p = 18.0$\,mag) exhibits 15 outbursts in 78.7 days data from \emph{K2} Campaign 8.  These recur on an average timescale of 4.9 days, reach 6--9\% peak fluxes, and last for 6--13 hours.

The full \emph{K2} light curves of these new targets are displayed in the bottom two panels of Figure~\ref{fig:lcs}.  The light curves of the previously discovered outbursting DAVs are displayed for comparison in the upper four panels in order of their discovery.  Detected outbursts are highlighted with gray points in contrast with the black quiescent segments, and smoothed or long-cadence light curves are overplotted in red.

Figure~\ref{fig:fts} displays the Fourier transforms (FTs) of the full light curves (including outbursts) for all six known outbursting DAVs.  The 0.1\% false alarm probability (FAP) levels were calculated for the \emph{K2} targets using a bootstrapping approach \citep{Bell2015,Bell2016}. The significant pulsation modes detected in KIC\,4552982 \citep{Bell2015} are marked in the top panel. The FTs all show multiple wide bands of power in the range $\approx$700--1250 $\mu$Hz, corresponding to pulsation modes that wander in amplitude and frequency.  Most FTs also show more stable modes at > 1900 $\mu$Hz.  This dichotomy in mode behavior at low and high frequency seems to be a general feature of DAVs, as was presented by Stephen Fanale et al.\ and M.~H. Montgomery et al.\ at this workshop.

We have obtained spectra of these two new outbursting DAVs but have not uniformly reduced and analyzed them yet.  A full analysis of these new objects will be presented in an upcoming refereed publication.

\articlefigure{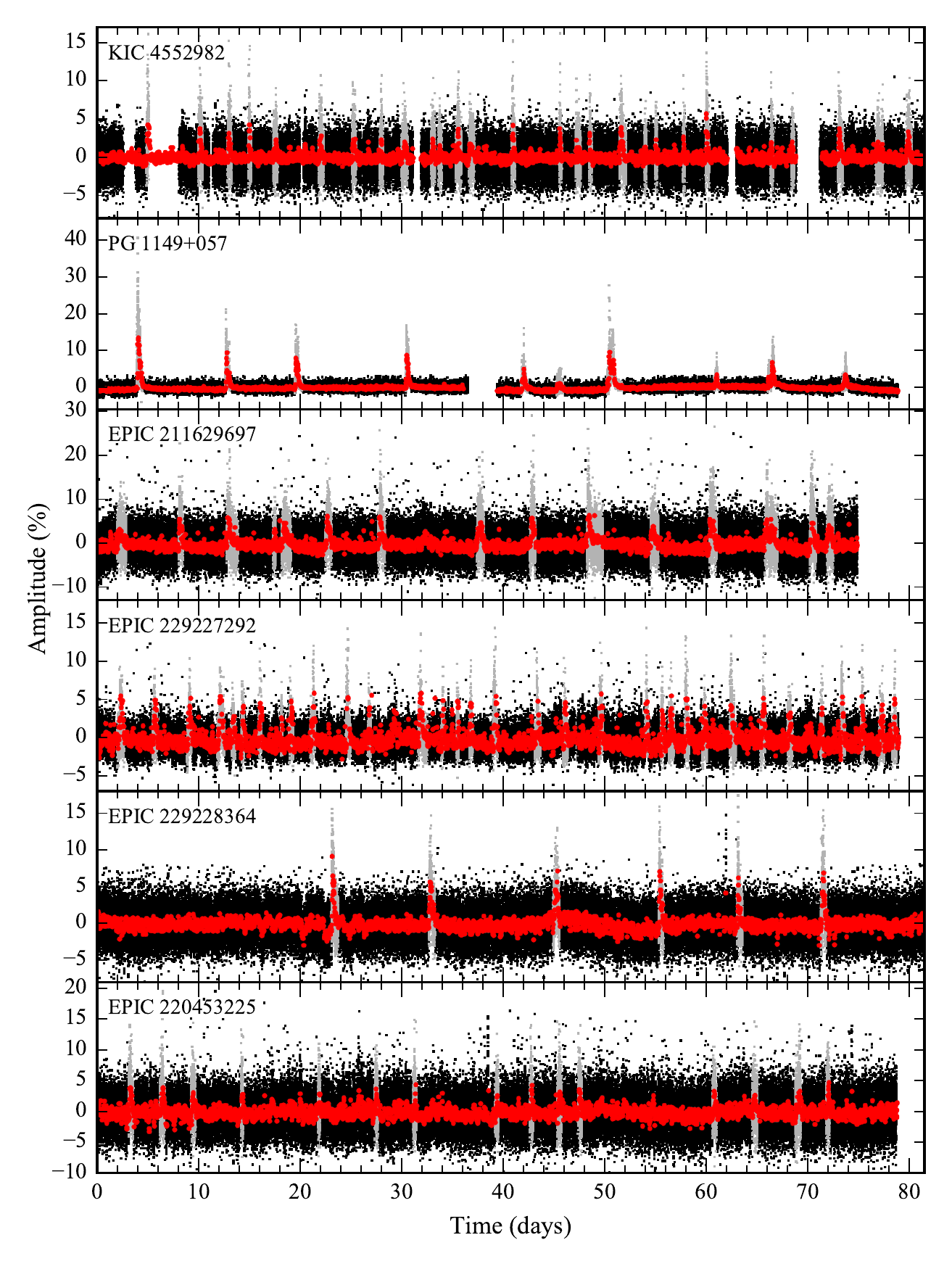}{fig:lcs}{The first 81.5 days of \emph{Kepler}/\emph{K2} observations of the 6 known outbursting DAVs, in chronological order of discovery from top to bottom.  The short-cadence (1\,min) data are displayed in black in quiescence and gray in outburst.  The long-cadence (30\,min) or 30-minute boxcar smoothed light curves are displayed in red. This view captures only a portion of the $>$1.5 year light curve of KIC\,4552982, but shows the light curves of the $K2$ targets in their entirety.  Each y-axis is scaled independently.}

\articlefigure{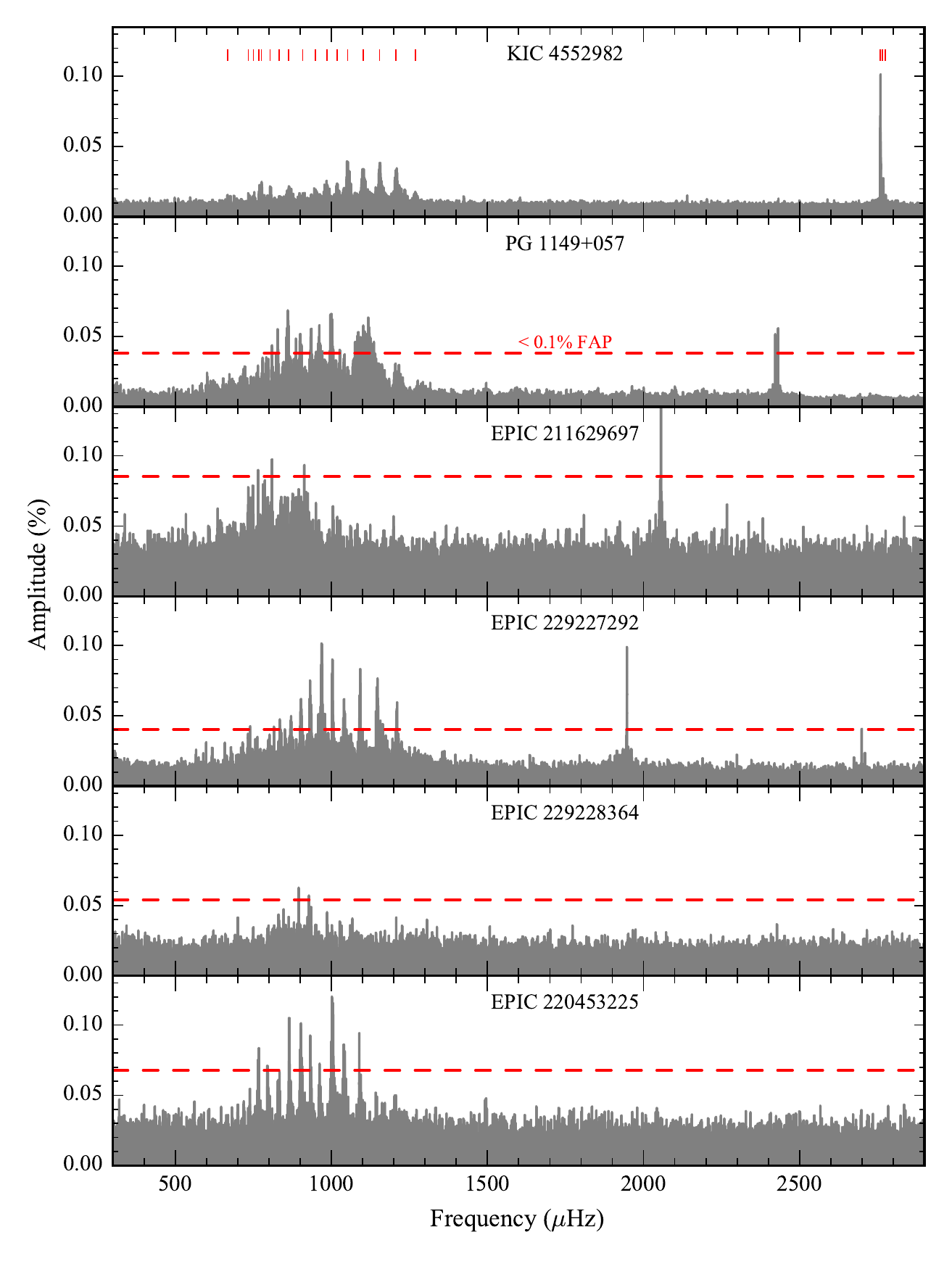}{fig:fts}{Fourier transforms (FTs) of the 6 known outbursting DAVs, in chronological order of discovery from top to bottom.  The first outbursting DAV, KIC\,4552982, is displayed with vertical tick marks at the locations of 20 significant pulsation frequencies determined by \citet{Bell2015}.  The latter 5 FTs include dashed significance thresholds where the false alarm probability (FAP) $<$ 1/1000 was determined from bootstrapping following \citet{Bell2016}. Each FT shows widened bands of power at frequencies $\lesssim 1250\ \mu$Hz, with more stable modes at higher frequency, as is typical of cool DAVs.}

\vspace{-.2in}

\section{The Class So Far}
\label{sec:class}

We summarize the observational characteristics of the six known outbursting cool DAVs from \emph{Kepler} in Table~\ref{tab:sum}.  We find generally that their properties are of the same order of magnitude across objects, but caution that many of these values can be affected by the signal-to-noise of targets of different magnitude in the \emph{Kepler} bandpass, so they may not be directly comparable between objects.

\vspace{-.1in}

\begin{table}[!ht]
\caption{Observational properties of outbursting DAVs \label{tab:sum}}
\begin{center}
{\small
\tabcolsep=0.085cm
\begin{tabular}{c c c c c c c c} 
\tableline
\noalign{\smallskip}
Name &$K_p$ & ${T}_{\mathrm{eff}}$	& $\log{g}$ & $\tau_{recur}$  & Med.\,Dur. & Max.\,Flux &Max.\,Energy\\
  & (mag) & (K)	&(cgs) & (d)  & (hr)  &  (\%) &  (erg) \\
\tableline\noalign{\smallskip}
KIC\,4552982\footnotemark[1]    & 17.9 & $10{,}860(120)$ & $8.16(0.06)$ & 2.7 & 9.6 & 17  & $2.1\times 10^{33}$  \\
PG\,1149+057\footnotemark[2]  & 15.0 & $11{,}060(170)$ & $8.06(0.05)$ & 8.0 &  15 & 45 & $1.2\times 10^{34}$ \\
EPIC\,211629697\footnotemark[3] & 18.4 & $10{,}570(120)$ & $7.92(0.07)$ & 5.0 & 16.3 & 15 & $1.8\times 10^{34}$ \\
EPIC\,229227292\footnotemark[3] & 16.7 & $11{,}190(170)$ & $8.02(0.05)$ & 2.4 & 10.2 & 9 & $3.1\times 10^{33}$ \\
EPIC\,229228364\footnotemark[4]  & 17.9 & $\ldots$ & $\ldots$ & 9.7 & 13.7 & 15 & $\ldots$\\
EPIC\,220453225\footnotemark[4]  & 18.0 & $\ldots$ & $\ldots$ & 4.9 & 8.6 & 9 & $\ldots$\\
\tableline
\end{tabular}
}

\end{center}
\end{table}

\footnotetext[1]{\citet{Bell2015}}
\footnotetext[2]{\citet{Hermes2015}}
\footnotetext[3]{\citet{Bell2016}}
\footnotetext[4]{This work}

\section{Looking Forward}
\label{sec:future}

Given the measured recurrence and duration timescales of these outbursts, capturing this behavior in time series photometry from a single ground-based observatory is a challenge. We cannot construct a precise ephemeris for these quasiperiodic events, and they will often occur while conditions are not favorable for observing.  As the light curve for EPIC\,229228364 in the fifth panel of Figure~\ref{fig:lcs} demonstrates, some outbursting DAVs can remain in a quiescent state for at least 23 continuous days.

Since outburst durations are often longer than a typical nightly run on a single object, terrestrial observers face an additional complication.  The flux of a target generally changes with wavelength across the observational bandpass differently than that of a comparison star, introducing to the divided light curve a signature from differential extinction as the telescope tracks through different parts of the atmosphere.  This is routinely addressed by dividing out low-order polynomials from DAV light curves during data reduction \citep[e.g.,][]{Nather1990}.  However, since we are now interested in astrophysical variations on $>$ hours timescales, this practice risks accidentally cleaning the light curves of outbursts.  A more sophisticated approach to mitigating the effects of differential extinction should be employed for this reason.  An effort to carefully reanalyze archival observations of DAVs for evidence of outbursts is currently underway.

\emph{K2} continues to visit new fields every $\approx$80 days, increasing the number of objects that we have such extensive data for.  Since outbursts seem to be a fairly common phenomenon in DAVs, we expect to continue to find new outbursters in upcoming data releases.

In the long term, the Large Synoptic Survey Telescope (LSST) promises to observe thousands of outbursting DAVs, each with at least $\approx$1000 visits across six photometric filters.  We anticipate that this survey will capture a fairly complete picture of outburst statistics and that the color information will reveal the temperature evolution of these events.  Real-time transient alerts from LSST will enable rapid photometric and spectroscopic follow-up of outbursts in progress.

\acknowledgements K.J.B., M.H.M., and D.E.W. acknowledge support from NSF grant AST-1312983, the Kepler Cycle 4 GO proposal 11-KEPLER11-0050, and NASA grant NNX13AC23G. Support for this work was provided by NASA through Hubble Fellowship grant \#HST-HF2-51357.001-A, awarded by the Space Telescope Science Institute, which is operated by the Association of Universities for Research in Astronomy, Incorporated, under NASA contract NAS5-26555. This work includes data collected by the \emph{Kepler} and \emph{K2} missions. Funding for the \emph{Kepler} and \emph{K2} missions is provided by the NASA Science Mission directorate.

\end{document}